# Improving MRI's slice selectivity in the presence of strong, metal-derived inhomogeneities

Gil Farkash, Gilad Liberman, Ricardo P. Martinho and Lucio Frydman*

*Department of Chemical and Biological Physics, Weizmann Institute of Science,*

*Rehovot 7610001, Israel*

*Prof. Lucio Frydman; +972-8-9344903; lucio.frydman@weizmann.ac.il

# Abstract

**Purpose:** To develop schemes that deliver faithful 2D slices near field heterogeneities of the kind arising from non-ferromagnetic metal implants, with reduced artifacts and shorter scan times.

**Methods:** An excitation scheme relying on cross-term spatio-temporal encoding (xSPEN) was used as basis for developing the new inhomogeneity-insensitive, slice-selective pulse scheme. The resulting Fully refOCUSED cross-term SPatiotemporal ENcoding (FOCUSED-xSPEN) approach involved four adiabatic sweeps. The method was evaluated *in silico, in vitro* and *in vivo* using mice models, and compared against a number of existing and of novel alternatives based on both conventional and swept RF pulses, including an analogous method based on LASER's selectivity spatial selectivity.

**Results:** Calculations and experiments confirmed that multi-sweep derivatives of xSPEN and LASER can deliver localized excitation profiles, centered at the intended positions and endowed with enhanced immunity to $B_0$ and $B_1$ distortions. This, however, is achieved at the expense of higher SAR than non-swept counterparts. Furthermore, single-shot FOCUSED-xSPEN and LASER profiles covered limited off-resonance ranges. This could be extended to bands covering arbitrary off-resonance values with uniform slice widths, by looping the experiments over a number of scans possessing suitable transmission and reception offsets.

**Conclusions:** A series of novel approaches were introduced to select slices near metals, delivering robustness against $B_o$ and $B_1^+$ field inhomogeneities.

**Introduction**

   Orthopaedic implants are widely used in modern healthcare (1,2). The biocompatible alloys used to manufacture these implants include stainless-steel, cobalt-chromium (CoCr) and titanium (3), with the latter preferred due to its small elastic modulus. The large base of prosthetic installations calls for methods to monitor the complications that often arise in the vicinity of implants; in particular osteolysis and aseptic loosing, conditions that are regularly asymptomatic prior to causing extensive bone loss (4). These complications go usually undetected by conventional X-rays, while computerized tomography (CT) exposes the patient to a significant dose of ionizing radiation and suffers from beam hardening artifacts near metallic devices (5). MRI can provide a solution to these challenges, particularly given that biocompatible alloys are at most only weakly ferromagnetic, and therefore not subject to strong forces or torques when placed in the static magnetic field. Still, metallic prosthesis will usually result in dramatic distortions in the images, with the main disruptive effects arising from metal-induced $B_0$ and $B_1$ perturbations. In a 3T $B_0$ field, for instance, tissue-metal susceptibility differences will result in off-resonance frequencies $\Delta f$ of up to $\pm 3$ kHz for titanium and up to $\pm 10$ kHz for stainless-steel (6). Under these conditions the application of a gradient will no longer arrange the spins' resonance frequencies linearly according to their spatial position, but rather will distribute them in a complex pattern determined by both the applied gradient and the spatial magnetic field disturbances $\Delta B_o = \Delta f/\gamma$ imparted by the implant. For a given 2D spatial density $\rho(x,y)$ within a thin body slice, signals will be influenced as

$$S(k_x, k_y) \propto \int_{slice} \rho(x,y) e^{-ik_x\left(\frac{2\pi\Delta f(x,y,z)}{\gamma G_{RO}}+x\right)} e^{-ik_y y} dxdy \qquad (1)$$

where $k_x$ and $k_y$ are the readout and phase encoding variables, $G_{RO}$ is a gradient applied during the readout that defines the in-plane extent of the distortion as $\Delta x = -\frac{2\pi\Delta f(x,y,z)}{\gamma G_{RO}}$, and the phase-encoded dimension $k_y$ is assumed unaffected thanks to its constant time nature. Also the slice selection, if performed by a frequency-selective RF (e.g., *sinc*) pulse applied in the presence of a magnetic field gradient $G_z$, will be distorted by the presence of these inhomogeneities. Given a pulse bandwidth *BW* attempting to excite spins within a slice $\Delta z$ centered at $z_o$, the frequency distortions will affect these positions in a complex (*x,y,z*)-dependence within $-\frac{BW}{2} < \frac{\gamma}{2\pi} G_z z +$ $\Delta f(x, y, z) < \frac{BW}{2}$ coordinates. This can lead to the excitation of a complex surface with variable thickness and varying *z*-centers across *x-y,* rather than of a planar slice. Distortions in these warped layers can be minimized by increasing the slice-selection bandwidth *BW*, thereby decreasing the $\Delta f(x, y, z)/BW$ spatial errors. The bandwidths, however, will be constrained by peak power limitations, or by limitations in the RF energy (Specific Absorption Rate, SAR) that can be deposited in a body.

Numerous proposals have been made over the years to deal with metal-driven field non-idealities. Although initially challenged by long acquisition times, multiple-coil arrays and under-sampling schemes have made full phase-encoding sequences viable routes to MRI free from static $B_0$ distortions (7, 8). A conceptually different approach involves view angle tilting (VAT), which modifies 2D spin-echo (SE) or turbo-spin-echo sequences by applying throughout their acquisitions a gradient equal the one used over the slice-selective excitation (9). Slice-selection errors near metal implants have also been addressed by techniques like Multi-Acquisition Variable Resonance Image Combination (MAVRIC) (11) and Slice Encoding for Metal Artifact Correction (SEMAC) (12). In MAVRIC slice selection errors are avoided by using a non-selective excitation, followed by 3D MRI using a readout and two phase encodings. A "spectral" dimension is then encoded in MAVRIC by stepping the transmission and reception offsets, leading to binned images which when subsequently summed, deliver results with remarkable robustness against $B_0$ inhomogeneities (11). By contrast SEMAC applies slice-selective excitation pulses but relies on their phase-encoding, so as to position the regions that were selectively excited in their correct *z* positions within the neighbourhood of the metal (12). Although MAVRIC and SEMAC treat inhomogeneities differently, their underlying physical principles are somewhat analogous: both techniques include an additional phase encoding dimension as well as a multiple-offset excitation, leading to extended scan times compared to regular 3D MRI. Common as well is their need to deal with folding artefacts along the phase encoded dimensions; the MAVRIC-SL variant addresses this issue along *z* by incorporating a weak slice-selective gradient during excitation and signal acquisition that selects only relevant slices prior to 3D *k*-imaging –even if at the cost of some blurring (13–15).

The present study explores slice-selectivity optimizations based on the recently introduced cross-term SPatio-Temporal ENcoding (xSPEN) strategy (10). xSPEN was proposed as a single-shot imaging technique that eliminates in-plane $B_0$ distortions, by applying a pair of identical swept pulses in conjunction with a constant $G_z$ and with bipolar $\pm G_y$ encoding gradients. Spins inside the targeted ROI are thus endowed with a bilinear phase term *Cyz*; the stationary point of this phase term can be driven from one side of the *y-axis* ROI to the other using a $G_z$ gradient, leading to a signal that is proportional to the spins density $\rho(y)$, regardless of the in-plane $B_0$ inhomogeneities (10). This study explores the potential of these concepts for achieving slice selectivity, and thus to facilitate the conventional acquisition of minimally-distorted 2D images in the vicinity of metals. In its original form, however, xSPEN –like other spatiotemporally encoded counterparts (19–25)– relied on frequency-swept pulses making all its manipulations progressive; as such it will impose a spatial dephasing that is not ideally suited for performing a conventional 2D acquisition. To deal with this Fully refOCUSED (FOCUSED) xSPEN derivatives are developed, and their ability to perform 2D slice selections in the presence of $B_0$ and $B_1$ distributions is evaluated. For completion the study also discusses LASER-derived options that, also based on swept pulses, can deliver faithful 2D slice excitations in the presence of field inhomogeneities. Evaluations involving numerical, phantom and *in vivo* comparisons among these as well as other slice-selection alternatives, are presented and discussed.

**Materials and Methods**

*Theoretical Considerations.* Figure 1a introduces the FOCUSED-xSPEN pulse sequence. The sequence was designed based on xSPEN's use of two 180° frequency swept pulses in the presence of a constant gradient and of a bipolar one (10, 26, 27); when considering that in the present instance spins will be subject to the constant effect of a sizable field inhomogeneity $\Delta B_o = \Delta f/\gamma$, the need for applying a constant external gradient disappears: $\Delta B_o$ will take this role, and the encoding part of the sequence becomes the segment encased by the dashed rectangle in Figure 1a. The sequence's adiabatic RF inversion pulses linearly sweep a bandwidth BW = |$O_i$ - $O_f$|, where $O_i$ and $O_f$ are initial and final RF offsets; these 180° pulses are executed in the presence of alternating $\pm G_z$ gradients defining the slice-selection dimension, and act after a pulse exciting the entire slab has been applied. Since our primary objective is to eliminate slice-selection errors induced by off-resonances, it is useful to examine the selectivity of the resulting block in a 2D space comprised by the resonance-offset ($\Delta f$) and the slice-select ($z$) dimensions. The first refocusing pulse, coupled to a negative $-G_z$ gradient, sweeps through a diagonal slab in $\Delta f$-$z$ defined according to $-\frac{BW}{2} + \frac{\gamma}{2\pi} G_z z < \Delta f(x, y, z) < \frac{BW}{2} + \frac{\gamma}{2\pi} G_z z$. Spins will accrue a spatially dependent quadratic phase inside this region, which at the conclusion of the first adiabatic pulse will be

$$\varphi^{(1)}(z) = -\frac{(\gamma G_z)^2}{R} z^2 - \gamma G_z \left( T_p + \frac{2(O_i - \Delta f)}{R} \right) z + \left( \Delta f \cdot T_p - \frac{(O_i - \Delta f)^2}{R} \right) \quad (2)$$

where R is the pulse's sweeping rate and $T_p$ is the swept pulse duration (21). Figure S1a (Supporting Information) shows the resulting phase in the 2D $\Delta f$-$z$ plane. The second refocusing pulse, coupled to a positive gradient $+G_z$, sweeps through a spectral-spatial region $-\frac{BW}{2} - \frac{\gamma}{2\pi} G_z z < \Delta f(x, y, z) < \frac{BW}{2} - \frac{\gamma}{2\pi} G_z z$. This also subtends a diagonal in $\Delta f$ and $z$ that this time it is aligned in the opposite $z$-direction vs the first slab –thereby defining a diamond in ($\Delta f$,$z$). The phase added by this second swept pulse differs from Equation 2 only by the sign of $G_z$, which has now been reversed. Taking the difference between two such terms, the overall evolution phase imparted on the spins as a result of these two manipulations is given by

$$\varphi^{(2)}(z) = 2\gamma G_z \left( T_p + \frac{2(O_i - \Delta f)}{R} \right) z = -\frac{4\gamma G_z}{R} \Delta f \cdot z + 2\gamma G_z \left( T_p + \frac{2O_i}{R} \right) z \quad (3)$$

Supporting Information Fig. S1b illustrates the ensuing phase profile, which contains a linear *z*-term and a bilinear *Δf·z* term that are detrimental for conventional (though not for xSPEN-based) imaging, and need to be refocused. One way of eliminating these is by adding a short non-selective π pulse, followed by a pair of swept pulses identical to the first pair. Another alternative, chosen here due to better robustness and presented in Figure 1a, concatenates an additional pair of frequency swept pulses that are similar to the first pair, except for their reversed sweeping direction. Due to their opposite sweep directions, these swept pulses will add to the spins an evolution phase term

$$\varphi^{(3)}(z) = \frac{4\gamma G_z}{R}\Delta f \cdot z + 2\gamma G_z\left(T_p - \frac{2O_i}{R}\right)z, \qquad (4)$$

leaving a linear phase term $4\gamma G_z T_r z$, which can be refocused using an appropriate rephasing gradient lobe (Supporting Information Fig. S1d). As a consequence of all these manipulations the diamond-shaped region selected by FOCUSED-xSPEN in *Δf-z* (Fig. 2a) will have no accrued phase. Notice that the selected region will narrow along the off-resonance dimension *Δf* (Fig. 2a), reducing the effective slice thickness as the off-resonance increases or decreases from the central frequency of the sweeps. Notice as well that despite this narrowing, and assuming that the initial excitation pulse is sufficiently broad-banded, the off-resonance range of the excited spins will always be between -BW/2 and +BW/2, and it will always remain centered along the *z* axis. The 2D diamond thus imparted by this procedure is analogous to that of xSPEN; except that in xSPEN MRI the diamond is defined by two spatial (e.g., *y-z*) dimensions, while in FOCUSED-xSPEN the selection occurs in the *Δf-z* space.

While the position of the desired *z*-slice in FOCUSED-xSPEN is accurately defined, the resulting approach is still narrowband in the sense of addressing up to ±BW/2 offsets. If BW/2 or if the initial excitation bandwidth are smaller than the *a priori* unknown inhomogeneity, part of the object at this particular *z*-slice will not be addressed. This can be overcome by looping the overall selection block (Figure 1a) over multiple transmission and reception offsets in steps *Δf*<sub>trans</sub>. Several frequency-binned images will then arise, corresponding to shifted *Δf-z* diamonds along the off-resonance dimension. These shifted diamonds will overlap or not depending on $\Delta f_{trans}$; as shown in the center row of Fig. 2a, setting this increment to $\Delta f_{trans}=BW/4$ will make diamonds overlap with their upper and lower neighbors along $\Delta f$. Summing all these images will thus yield a "slice" centered in $z$, covering the off-resonance dimension with a uniform response and a well-defined z-thickness. Note that for all these off-centered $\Delta f$-$z$ diamonds the signal emitted has to be recorded with a reception offset that is also shifted by $\Delta f_{trans}$; this is akin to the requirement arising in MAVRIC (11).

By relying on adiabatically-swept pulses, FOCUSED-xSPEN raises associations with Localization by Adiabatic SElective Refocusing (LASER) (19), a sequence that relies on pairs of adiabatic pulses for selecting a given region (Figure 1b). LASER and its derivative semi-LASER are well-known localization sequences for MRS due to their robustness against $B_0$ and $B_1$ inhomogeneities (19, 28–30). When considered in the context of a sizable inhomogeneity $\Delta f$, LASER pulses executed with a bandwidth $BW$ will lead to a spatially bound parallelogram, tilted in $\Delta f$-$z$ space (Fig. 2b). To reinstate constant response across $\Delta f$ we extend the approach used for the xSPEN pulses and propose implementing Frequency-Offset-Correction USing iteratED LASER (FOCUSED-LASER), where the LASER block is looped in multiple shots over transmitter and receiver offsets separated by $\Delta f_{trans}$. This will lead to a chain of spectrally shifted $\Delta f$-$z$ parallelograms (Figure 2b, center row), which will overlap if $\Delta f_{trans}$ is sufficiently small (e.g., $\leq \frac{BW}{2}$). Notice that if $\Delta f_{trans}=BW/4$ these parallelograms will be nearly overlapping (Fig. 2b), but an effective $\Delta z$ thickness larger than originally intended, will arise. Notice as well that for equal bandwidths and durations of the swept pulses FOCUSED-xSPEN will deposit twice as much energy as FOCUSED-LASER; to account for this, the slice-selection bandwidths of the swept pulses used here in the FOCUSED-LASER comparisons were doubled when assessing their performance. The bottom row in Fig. 2b complements these considerations by showing this sequence's robustness against $B_1^+$ heterogeneities.

In a recent elegant development Hargreaves *et al* (31–33) introduced a technique that is related to the FOCUSED-xSPEN version described above: 2D Multi-Spectral Imaging (MSI, Fig. 1c). 2D MSI uses slice-selective Shinnar-Le Roux (SLR) pulses for excitation and refocusing in the presence of negative (-$G_z$) and positive (+$G_z$) slice-selection gradients, respectively. Like in xSPEN, the combined action of these two selective pulses selects a spectral-spatial diamond in *Δf-z* (Fig. 2c), which when looped over transmission offset shifts *Δf_trans_* extends the selectivity into a train of spectral-spatial diamonds. Clearly this is the simplest and lowest SAR of all the variants, and as such it is included in the analysis in Fig. 2.

*Numerical and Experimental* All simulations (Figure 2, also Supporting Information) were calculated in Matlab (The Mathworks, Natick, MA) as the real or absolute values of the Fourier Transform (FT) for each readout signal; experimental images were also processed with Matlab scripts, available upon request. Experiments tested the performance of FOCUSED-xSPEN against LASER and FOCUSED-LASER (Figure 1); for completion comparisons also included a Spin-Echo (SE) sequence. To achieve conditions of equal SAR, the frequency swept pulses in the LASER and FOCUSED-LASER experiments were given twice the bandwidths as their FOCUSED-xSPEN counterparts; naturally, the SE deposits a much lower SAR.

All experiments were performed on a DD2® 7T/110mm horizontal magnet scanner (Agilent Technologies, Santa Clara, CA) equipped with a quadrature birdcage volume coil probe. All experiments were conducted through the Agilent VNMRJ® imaging software, running custom-built pulse sequences available upon request. Since off-resonances caused by susceptibility distortions scale linearly with magnetic field and RF non-uniformities generally intensify at higher fields, image degradation under these conditions are probably stronger than for the same phantoms scanned at 3T.

One of the phantoms used in the *ex vivo* experiments consisted of a tube filled with $CuSO_4$-doped water, containing a Lego piece and a titanium screw. This screw was collinear with the tube's long axis and attached to the Lego; the tube was positioned in the center of the magnet and aligned parallel to the static magnetic field. In this configuration, pulse-acquire experiments showed a line-width of approximately 5 kHz. Percentage image uniformity (PIU) values near the metal were calculated according to the American College of Radiology (ACR) standard (34) as $PIU = 100 \cdot \left[1 - \frac{S_{max}-S_{min}}{S_{max}+S_{min}}\right]$, where $S_{max}$ and $S_{min}$ correspond to the maximal and minimal pixel intensities in a ROI near the metal comprised of 20 x 20 pixels and placed over a homogenous water region. Additional phantoms were built from a cucumber with a titanium screw inserted in its center and placed parallel to field, leading to a similar line width.

*In vivo* experiments were preapproved by Weizmann's Animal Care and Use Committee (protocol 01500312-3), and were carried in accordance with the guidelines of this IACUC. These tests included taping titanium disks (10mm diameter, 2mm thickness) in-between the ears of mice, and recording head images from the animals. During their course a mixture of isoflurane (1-2%) and oxygen was provided through a dedicated nose mask; this anesthesia was flowed at $\approx$1 L/min and the animal's respiration was monitored throughout from a suitable sensor (SA Instruments). The animal was placed on top of a laboratory-built system circulating warm water, maintaining a stable body temperature. Experiments were performed on different days, scanning different mice.

The sequences assayed were as schematized in Figure 1, and they were performed in a multi-shot framework iterating over transmission/reception offsets and over phase encoding steps. All xSPEN and LASER sequences began with an excitation pulse with bandwidth of 4 kHz; WURST-shaped pulses (35) were used for the linear chirps; unfortunately, the MSI sequence was not available in our scanner and thus could not be suitably compared with other methods. Similar *k*-acquisition schemes were applied for all 2D MR methods, collecting 64 $k_y$ phase encoding steps and 128 $k_x$ readout samples with an acquisition bandwidth of 250 kHz, resulting in an acquisition time of 0.5 ms for each line in *k*-space. Projection images in Figure 4 were recorded by applying a readout gradient as that illustrated in the left-hand side of Figures 1a, 1b (i.e., along the *z* slice-selection axis), in order to reveal the sculpting accuracy of each slice-selective technique. Since the *x* dimension was not spatially encoded in these experiments these results become projection images, in which all signal contributions along *x* appear integrated.

## Results

Figure 3 displays *x-y* images arising from various slice-selective techniques when applied on a phantom containing a titanium screw, for comparable signal acquisition conditions. Selected regions for certain experiments (blue squares in Figure 3) are reproduced, zoomed, in Supporting

Figure S2 for ease of comparison. The largest metal-induced distortions arise when applying multi-shot SE MRI in a conventional, on-resonance fashion (Fig. 3d). LASER applied with a doubled slice-selection bandwidth (Fig. 3c) shows a remarkable improvement over SE, but the shading surrounding the screw is still larger than what is seen in FOCUSED-xSPEN (Fig. 3a). By looping over a number of $\Delta f_{trans}=BW_e/4$ offsets, FOCUSED-LASER (also applied in a doubled slice selection bandwidth vis-à-vis FOCUSED-xSPEN) is somewhat better than LASER in reducing the shading next to the metal (Fig. 3b) –even if its details are still poorer than those seen in Figure 3a. "Stitching" problems may also be conspiring against these sequence's results. Uniformity values measured for $Z=0.4$ cm (dashed red squares in Fig. 3) performed as described in Methods, gave 79%, 70%, 10% and 4% for the FOCUSED-xSPEN, FOCUSED-LASER, LASER and SE experiments, respectively. These values changed somewhat upon moving the chosen ROI chosen but reflect the overall trend revealed by the images; see for instance the discussion presented in connection to Supporting Figure S3. Supporting Figure S4 plots 1D profiles extracting from these phantom experiments, to illustrate how the various approaches preserve the sharpness of a rapidly changing object (in this case, the Lego block). Additional results collected on the same phantom using a different positioning, can be found in Supporting Figure S5.

Figure 4 provides a different perspective on the slice-selection efficiency of these methods, by presenting profiles arising from the same phantom as in Fig. 3, but when imaged along the slice-selective axis. The slice profiles of multi-shot SE, applied without any transmission offsets, are again distorted. LASER profiles are better but still seem blurry, and FOCUSED-LASER only provides an incremental improvement on this. FOCUSED-xSPEN produces accurately localized slices that seem devoid of shading artifacts. Supporting Figure S6 provides additional information on these results, by plotting the 1D slice profiles arising from each technique along the slice-selective axis. These measurements also suggest that accurate, uniform slice profiles arise from FOCUSED-xSPEN. Yet another perspective concerning slice selectivity is presented in Supporting Information Figure S7, where slices were monitored using chemical shift imaging sequences relying on a full phase encoding of this spatial dimension. These measurements highlight the fact that main differences between the various sequences, relate not just to their different localizing abilities but rather to their handling of $B_1$ inhomogeneity effects.

Figure 5 presents another set of phantom results, this time recorded on a cucumber containing a titanium metal going through its center. Slices were selected here sagittaly, to better appreciate the effects of the metal on the object's fine structure. Once again, fewer shading artefacts arise for sequences relying on adiabatic sweeps rather than on monochromatic pulses, particularly in proximity to the metal. In these regions the quality of the FOCUSED-xSPEN exceeds that of LASER and FOCUSED-LASER –even if the latter appear to yield slightly better sensitivities for other regions. Additional (coronal) results collected on this phantom that further validate these conclusions, are presented in Supporting Information Figure S8.

Figure 6 compares the performance of various slice-selection schemes introduced in Fig. 1, when tested *in vivo* on a mouse on whose head a titanium disk has been placed (Figure 6, top panel). Field distortions induced by the metal ruin the integrity of the multi-shot SE images (Fig. 6a); LASER images (Fig. 6b) are remarkably less distorted than the SE counterparts, but still exhibit shading artifacts and additional distortions in the brain region (marked with yellow arrows). The FOCUSED-xSPEN results (Fig. 6c) exhibit the weakest shading artifacts, delivering faithful images from most brain areas.

Figure 7 shows a more extensive set of tests on a different animal, where the sweep bandwidths were doubled for LASER and for FOCUSED-LASER so as to account for their intrinsically lower SARs vs xSPEN. This brings about an improvement in the quality of the resulting images; still, even then, some of the perceived improvements actually reflect miss-registration effects arising from displacements from different slices (e.g., the bright region indicated by a yellow arrow in slice Z=-0.4 cm, arises from the hypointense region also indicated by an arrow in the Z=0 slice). Additional experiments done upon placing a smaller metal cylinder on top of a mouse's head, are shown in Supporting Information (Figure S9). In all cases, the FOCUSED-xSPEN approach consistently gave the weakest shading artifacts and most faithful images, even if in the immediate proximity of the metal disk the distortions dominated all methods assayed. Notice a certain loss of texture in the images even upon utilizing FOCUSED-xSPEN, probably reflecting the relatively large thicknesses (4 mm) used in these slice-selective mouse brain experiments (going to finer slices was challenged sensitivity-wise by the relatively large inhomogeneities involved in this instance, which compromised in turn the sensitivity arising from each frequency-incremented experiment).

## Discussion and Conclusions

Characterizing soft tissues proximate to metal devices is an important clinical problem, which MRI is uniquely well posed to solve. This, however, requires overcoming the large susceptibility-driven distortions arising in such instances –both inside the scanned planes as well as during the slice-selection process. Clinical approaches such as MAVRIC often correct for slice selection errors by adding a phase encoding loop in $k_z$; while general and powerful, this may prolong examination times and exhibit additional sensitivities to $B_1^+$ inhomogeneities arising from electrical currents induced in the metal and in its surrounding medium. The present study focused on designing slice-selection strategies that are relatively insensitive to both $B_0$ and $B_1$ inhomogeneities, by relying on frequency-stepped, chirped pulses. One such sequence exploited xSPEN's immunity to in-plane distortions caused by $B_0$ heterogeneities, to design an excitation scheme that delivers faithful 2D slices near implants with reduced scan times. To this effect the role usually taken by xSPEN's phase-encoded domain was replaced by the intrinsic field distortions arising along the slice selection axis. This could excite a well-defined slice for a range of offsets defined by the initial excitation and the subsequent chirped pulses' bandwidths; the large range of metal-derived offsets was then compensated by repeating the spatial excitation procedure across a series of off-resonance values. The potential of the resulting FOCUSED-xSPEN approach was verified with simulations, which confirmed that combining multiple frequency stepped acquisitions could deliver an accurate slice across a range of off-resonances. The slice would then be accurately centered on the prescribed $z$ position, its thickness would be uniform, and the intensities would be relatively insensitive to $B_1$ distortions. Given its reliance on multiple chirped pulses, FOCUSED-xSPEN also invited parallel comparisons with LASER, another well-known technique widely used in localized spectroscopy. Based on this FOCUSED-LASER, a derivative of LASER incorporating frequency-stepped loops concatenating the $\Delta f$-$z$ parallelograms associated with LASER's selection along the off-resonance dimension, was devised. Once again, simulations predicted attractive features for this form of slice selection. Notice, however, that simulations are not necessarily reflective of sensitivity constraints: even if bypassed by improved excitation schemes, large inhomogeneities may lead to spatially restricted excitations, where the number of contributing spins is limited and hence SNR insufficient. This constraint is likely to lead to limits in the spatial localization afforded by any band-limited approach, regardless of its theoretical merits.

Still, despite dealing with intra-slice off-resonances exceeding ±3.5 kHz, experiments carried out at 7T showed promising results for these new sequences both in phantoms and *in vivo*: FOCUSED-xSPEN systematically evidenced the least shading and missregistration artifacts, among a range of techniques collected under comparable acquisition times, spatial resolutions and FOVs. This was true even when comparing FOCUSED-xSPEN to LASER or FOCUSED-LASER experiments possessing twice the slice selection bandwidths –reflecting the different physics of the two experiments. All methods based on adiabatically swept RF pulses, showed comparable robustness to RF inhomogeneities. 2D MSI brings a clear advantage over other methods in terms of combining a partial compensation for inhomogeneities, with a much lower SAR per scan than any of the frequency-swept-based sequences. However, 2D MSI could be more affected by RF inhomogeneities, as simulations suggest. This issue will be explored on clinical scanners in future studies; robustness to RF inhomogeneities could prove decisively advantageous in instances were higher powers are associated to local metal-induced heating effects.

Ongoing investigations are exploring whether the looped excitations of multiple diamonds or parallelograms in $\Delta f$ and z introduced for the FOCUSED sequences, are really necessary. Calculations show that by adopting multiband or PINS principles along the $\Delta f$ domain (36, 37), an entire span of off-resonances can be excited in a single shot, thereby eliminating the need to iterate scans as function of transmission offsets. Discretized slice-selective RF pulses could also be used to excite simultaneously several diamonds in the full 2D $\Delta f$-$z$ domain; as in such case diamonds belonging to different slices would be endowed with unique linear phase terms (18, 38), shifting their echoes along the *k-* readout and enabling separation of echoes from different slices would be an added plus. Also worth pursuing is the combination of these slice-selective approaches with robust single-shot diffusion measurement techniques –as provided by xSPEN single-shot MRI (39). These methodological developments, as well as extensions to human clinical studies, are in progress.

**Acknowledgments.** We are grateful to Dr. Qingjia Bao (WIS) for help in handling the animals, and to Drs. Jean-Nicolas Dumez (CNRS, France), Amir Seginer (Siemens Healthineers Israel) and Zhiyong Zhang (WIS) for valuable discussions. This research was supported by ISF grant 965/18, ISF/NSFC grant 2508/17, the Kimmel Institute for Magnetic Resonance (WIS), and the generosity of the Perlman Family Foundation.

**Supporting Information.** Additional information and examples can be found in the Supporting Information section available in the on-line version of this article.

**Figures**

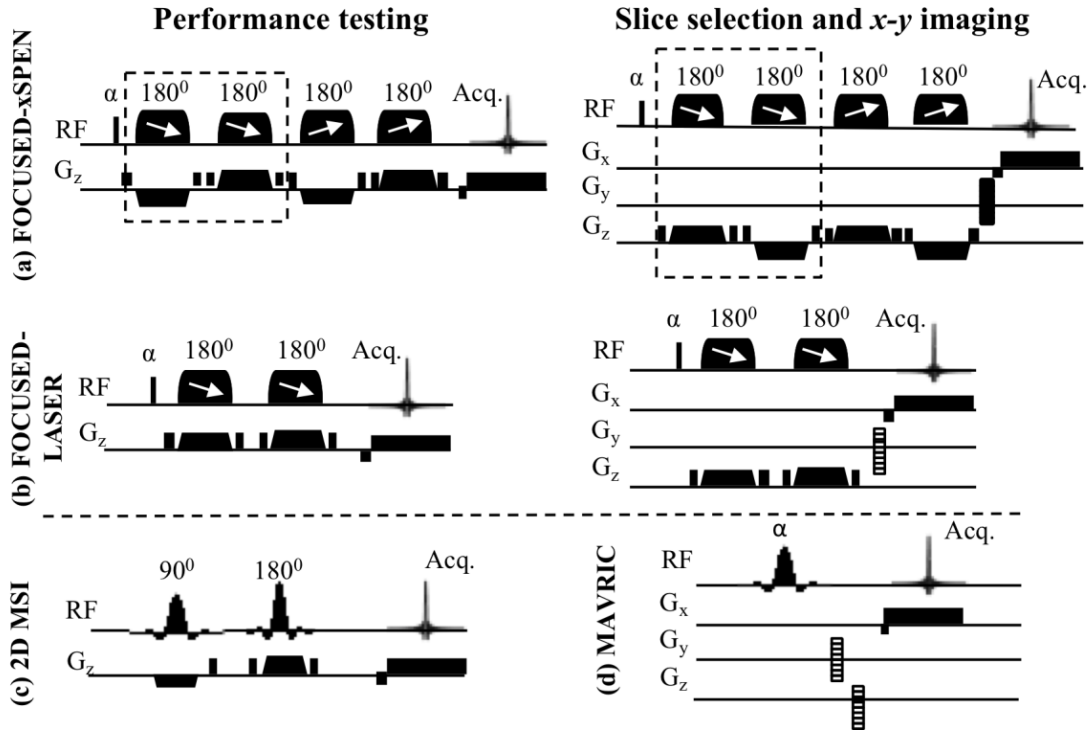

**Figure 1:** Summary of various slice-selective approaches examined or referenced in this study. (a) FOCUSED-xSPEN, highlighting in the dashed rectangle the original xSPEN hyperbolic-phase encoding element. (b) FOCUSED-LASER, a new sequence proposed as a simpler FOCUSED-xSPEN alternative. (c) 2D MSI, a bipolar version of multi-scan SE (32), applying *sinc* RF pulses coupled to alternating $G_z$ gradients for improving slice localization. (d) MAVRIC sequence (11) involving a frequency selective pulse and subsequent 3D *k*-imaging with two phase encodings in *y,z* and a readout in *x*. Sequences (a)-(b) are shown on the left column as executed for testing their performance, which included a single imaging readout acquisition along the slice-selective axis in order to better illustrate the slice selective. Shown on the right-hand column are examples of the final role intended for such sequences for facilitating a slice-selective excitation along *z* and conventional *k*-imaging in the x-y plane; notice that in these instances no particular provision was included for the sake of speeding up any of the acquisitions.

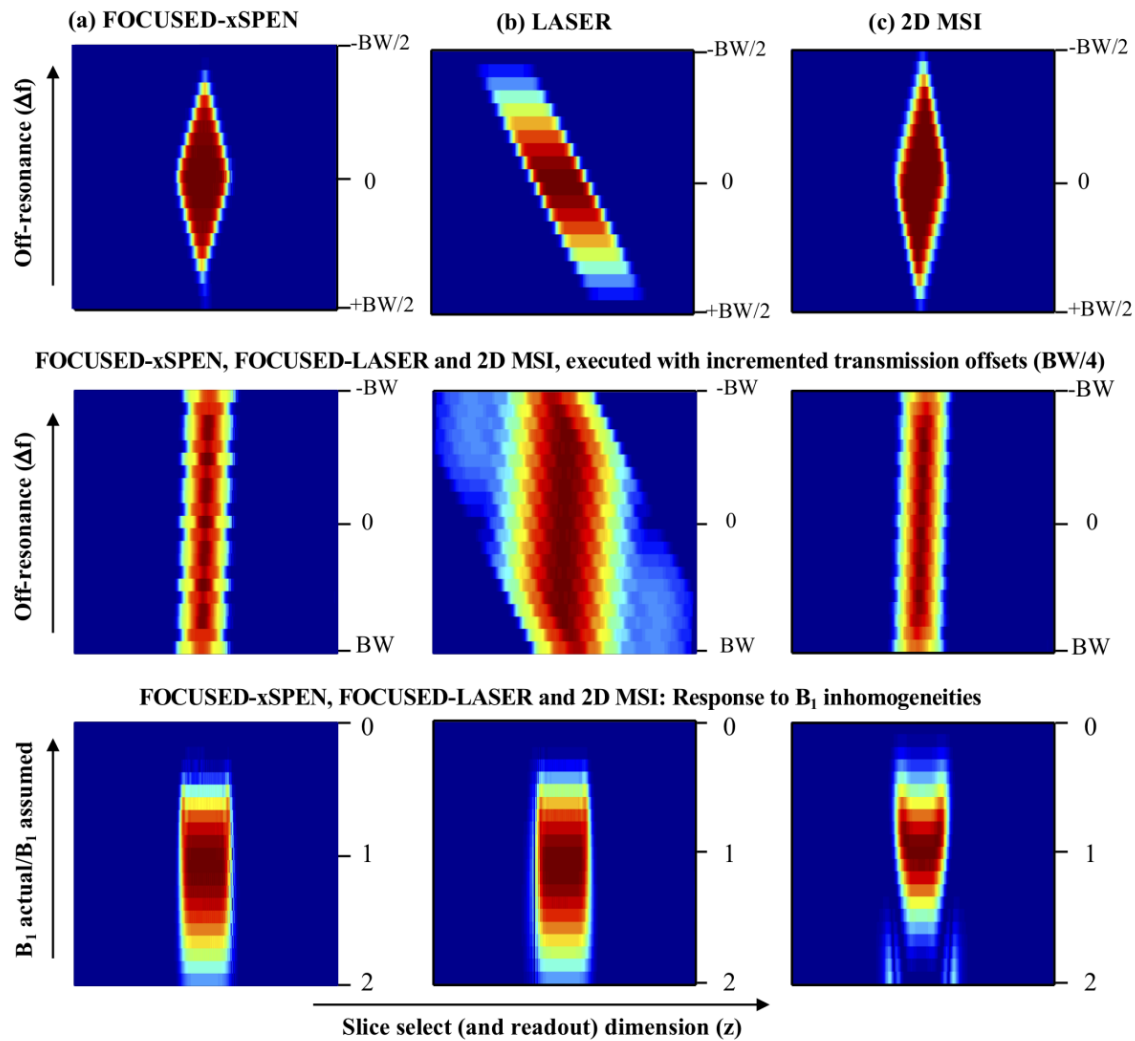

**Figure 2:** Regions selected by various sequences in the offset (*Δf*) / spatial (*z*) dimensions, illustrated for magnitude-mode images arising from Bloch simulations of the sequences in the left-hand column of Figure 1. In all cases simulations were done by considering 20,000 discrete spatial elements, and propagating the spin-evolution of these elements throughout the excitation, encoding and data acquisition processes in steps (dwell times) of 4 µs. (a) FOCUSED-xSPEN based on two pairs of swept pulses addressing a bandwidth BW. Top row: Diamond selected by a single xSPEN block in *Δf* (vertical) / z (horizontal). Center row: Results arising upon repeating this block with different transmission offsets, incremented by steps of BW/4; notice the uniform response across *Δf*. (b) Idem but for a looped LASER (FOCUSED-LASER) pulse scheme, with the center frequency offset steps set equal to the excitation BW/4. (c) Idem but for 2D MSI, with the looped version arising from transmission and reception offsets increments in steps of BW/4. For this row images are sums of all frequency-binned experiments. Bottom row: sensitivity of each experiment to $B_1^+$ inhomogeneities covering an RF magnitude scaling factor ranging from 0 to 2 –representative of reports at 3T for phantoms containing a titanium rod (39). Frequency swept pulses in FOCUSED-xSPEN, LASER

and 2D MSI had a 4 kHz bandwidth, associated with a slice thickness of 4 mm. All acquisitions monitored a 2 cm FOV and collected 256 readout samples, corresponding to a pixel size of 0.078 mm, in an acquisition time $T_a=1$ ms. For the panels in the upper row, the scan time for each 1D image would have taken 23 ms for (a) and 16 ms for (b-c).

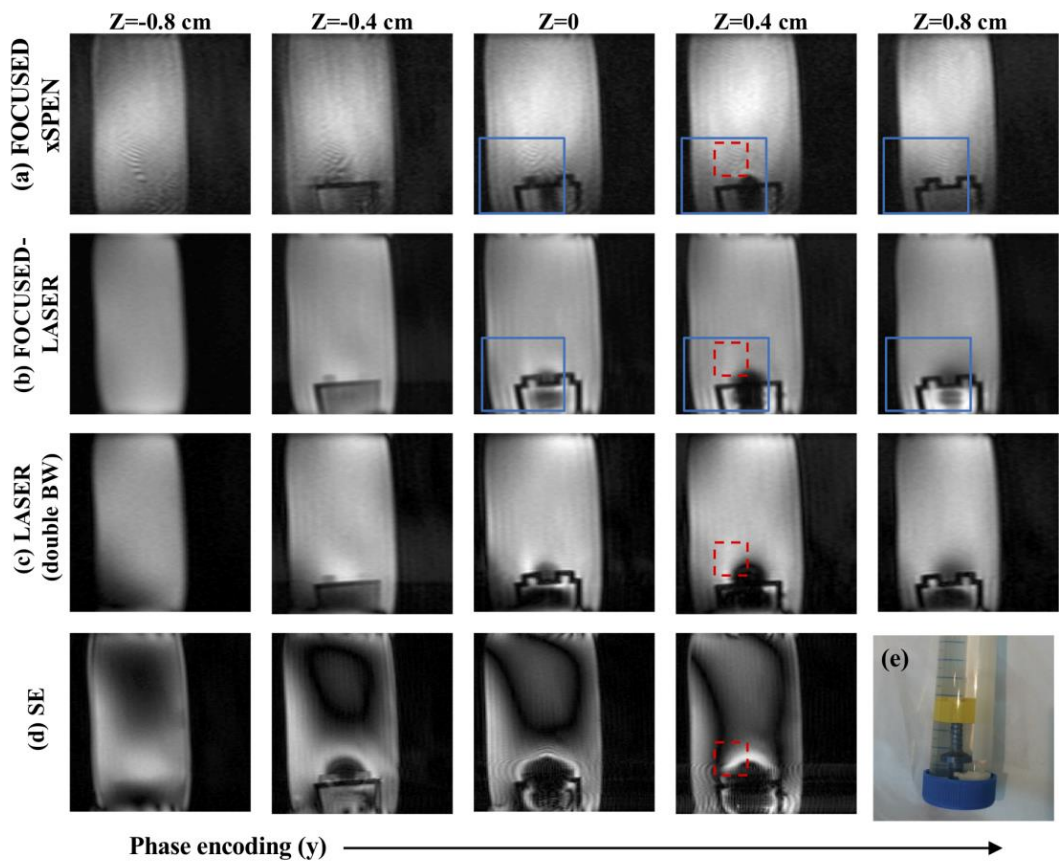

**Figure 3**: Spin-echo images of a $CuSO_4$-doped water tube containing a Lego piece and a titanium screw (e), illustrating the performance of various slice-selective approaches. (a) FOCUSED-xSPEN slice selection arising from pairs of BW = 4 kHz, 6 ms long swept pulses, covering a slice thickness of 4 mm. (b) FOCUSED-LASER images acquired using the sequence in Fig.1b (right-hand column), while looping over spectral shifts with increments of 1 kHz, using twice the sweep bandwidths as in (a). (c) Idem but for LASER. (d) Multi-shot SE applied as (c), without incrementing transmission offsets. SE was executed using 4 ms/4 kHz *sinc* pulses. The loop over transmission and reception offsets consisted of 9 steps (a-b) with increments of 1 kHz. The repetition time (TR) was 0.4 s, while the echo times (TE) was 30 ms for (a), 20 ms for (b-c) and 11 ms for (d). The acquisition bandwidth was for all cases 250 kHz and 128 readout samples were collected, leading to a data sampling time of 0.5 ms. Total acquisition durations were: 15 min for (a-b), 1.6 min for (c) and 2 min for (d). The dashed red squares mark the regions used to calculate image uniformity; the full blue-line squares indicate zoomed regions extracted from the indicated images, and shown for further clarity in the Supporting Figure S2. See Supporting Information for this, as well as Figures S3-S5 for additional results obtained for different placements of the same phantom.

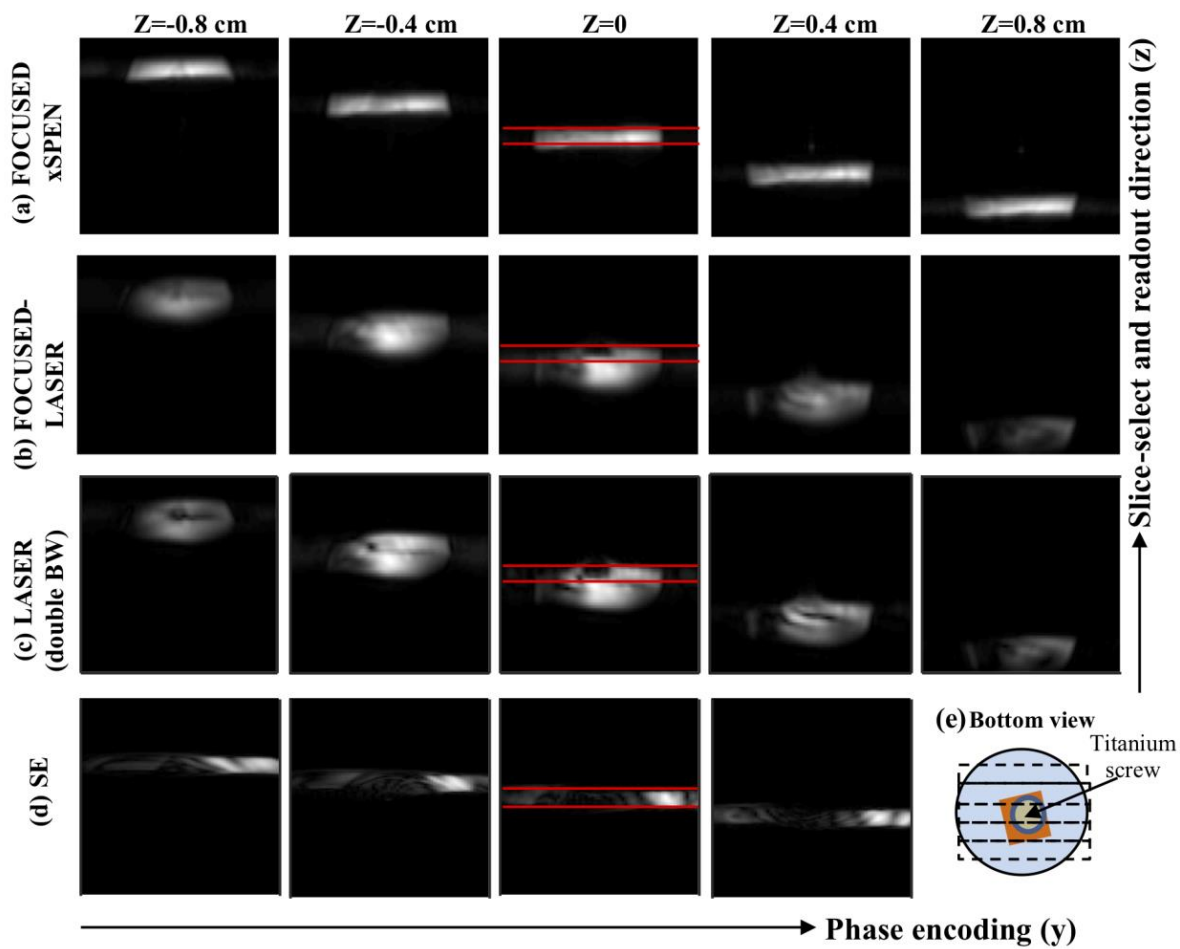

**Figure 4:** Comparing the slice selectivity accuracy of various sequences introduced in Fig. 1 ("Performance Testing" column), as evidenced upon applying the readout gradient along the slice-select dimension, while applying for all methods the same RF pulsing block as schematized in Fig. 1. The results are thus projected images of the titanium-containing phantom introduced in Fig. 3. The accuracy of the slice-selection is examined along several *z* offsets spaced 4 mm apart, with all sequences but LASER and SE collected with nine transmission offsets spanning 9 kHz. RF durations and bandwidth were identical to Fig. 3. Indicated with red bars are the slices targeted by the various experiments for Z=0. An alternative set of results relying on purely phase-encoded chemical shift imaging versions of these sequences on this phantom, is presented in the Supporting Information Figure S7.

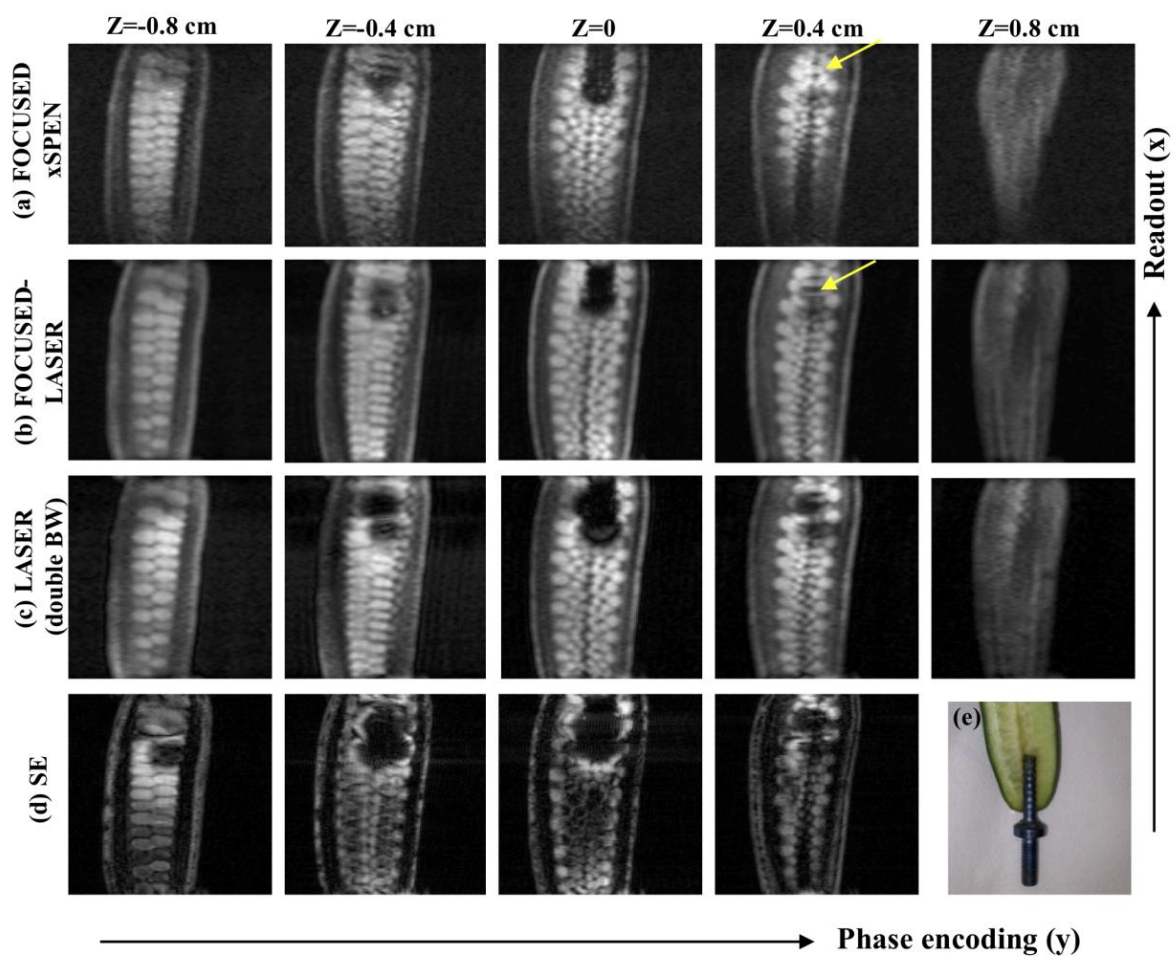

**Figure 5:** Sagittal images of a cucumber with an inserted titanium screw (sample shown in (e)) collected using: (a) FOCUSED-xSPEN; (b) FOCUSED-LASER without Δf overlaps; (c) LASER; (d) on-resonance SE. Sequence parameters were identical to Figure 2 in all aspects, meaning that LASER and FOCUSED-LASER swept a doubled bandwidth (BW = 8 kHz) compared to FOCUSED-xSPEN (BW = 4 kHz). Total acquisition times were 14.5 minutes for (a-b), 1.6 minutes for (c) and 2.1 minutes for (d). Slice thicknesses were set to 4mm. Yellow arrows indicate regions where subtle differences between the xSPEN- and LASER-based slice selectivities are visible. Notice the slightly different attenuations exhibited by the swept- vs hard-pulse images, probably as a consequence of the different echo times of these sequences. See Supporting Information Figure S8 for additional results on this phantom.

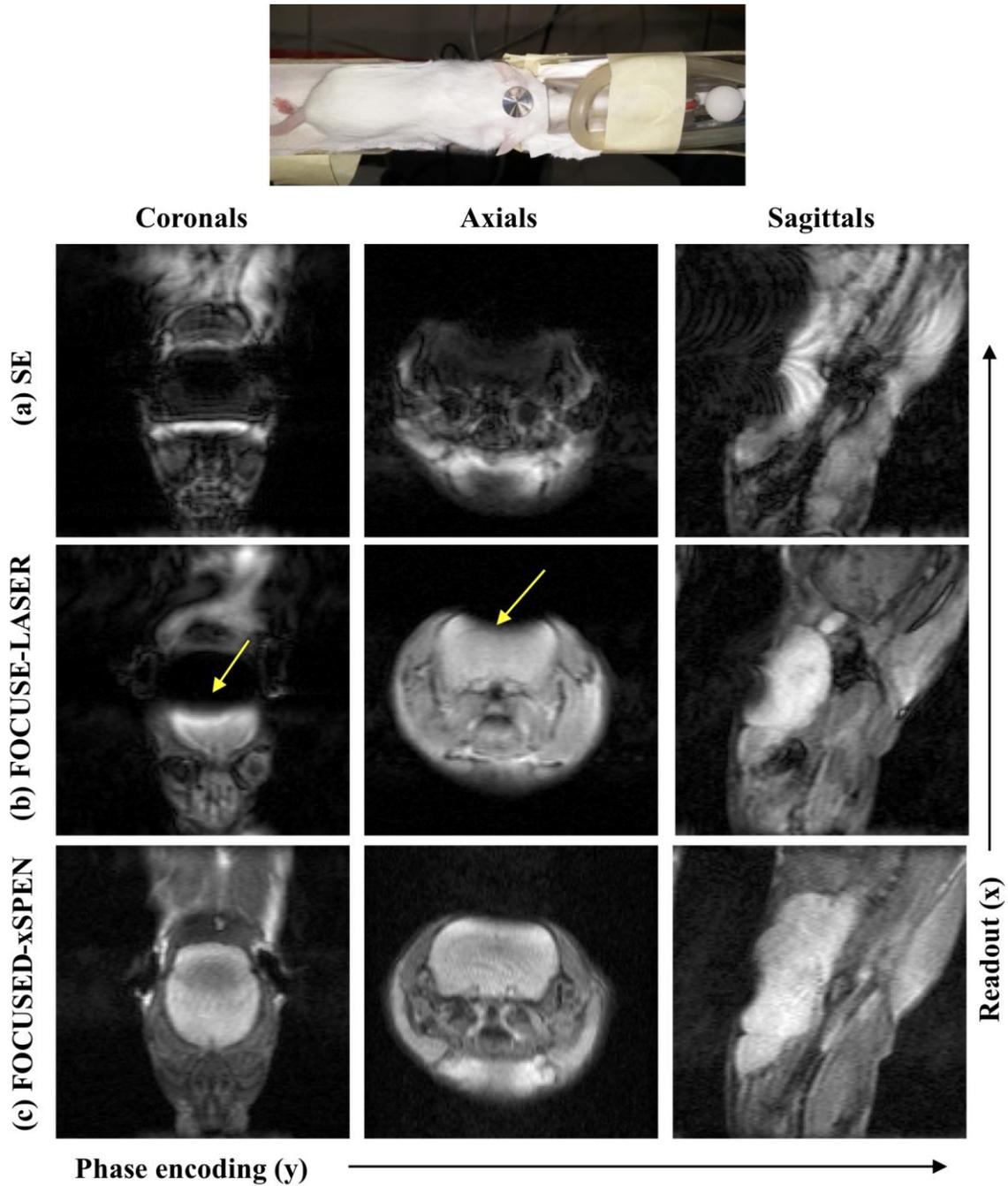

**Figure 6**. *In vivo* images of a mouse with a titanium disk placed on top of the skull (setup on top) acquired using: (a) a SE sequence; (b) FOCUSED LASER; (c) FOCUSED xSPEN. LASER and xSPEN images (experiments (a) and (b) in Fig. 1) were collected by frequency-stepping the pulses in 7 or 9 steps of 1 kHz respectively, and calculated as the quadrature sums of the binned images. Swept pulses in xSPEN and LASER were 4 ms long and covered a bandwidth of 4 kHz, associated with a slice thickness of 4 mm. Sinc pulses in SE were 4 ms long and affected a bandwidth of 4 kHz, while the LASER and xSPEN started with a 0.2 ms hard pulse, exciting a bandwidth of roughly 5 kHz (SAR-

wise terms this comparison is thus biased against FOCUSED-LASER). In all cases, 128 samples were collected during the readout, at an acquisition bandwidth of 250 kHz

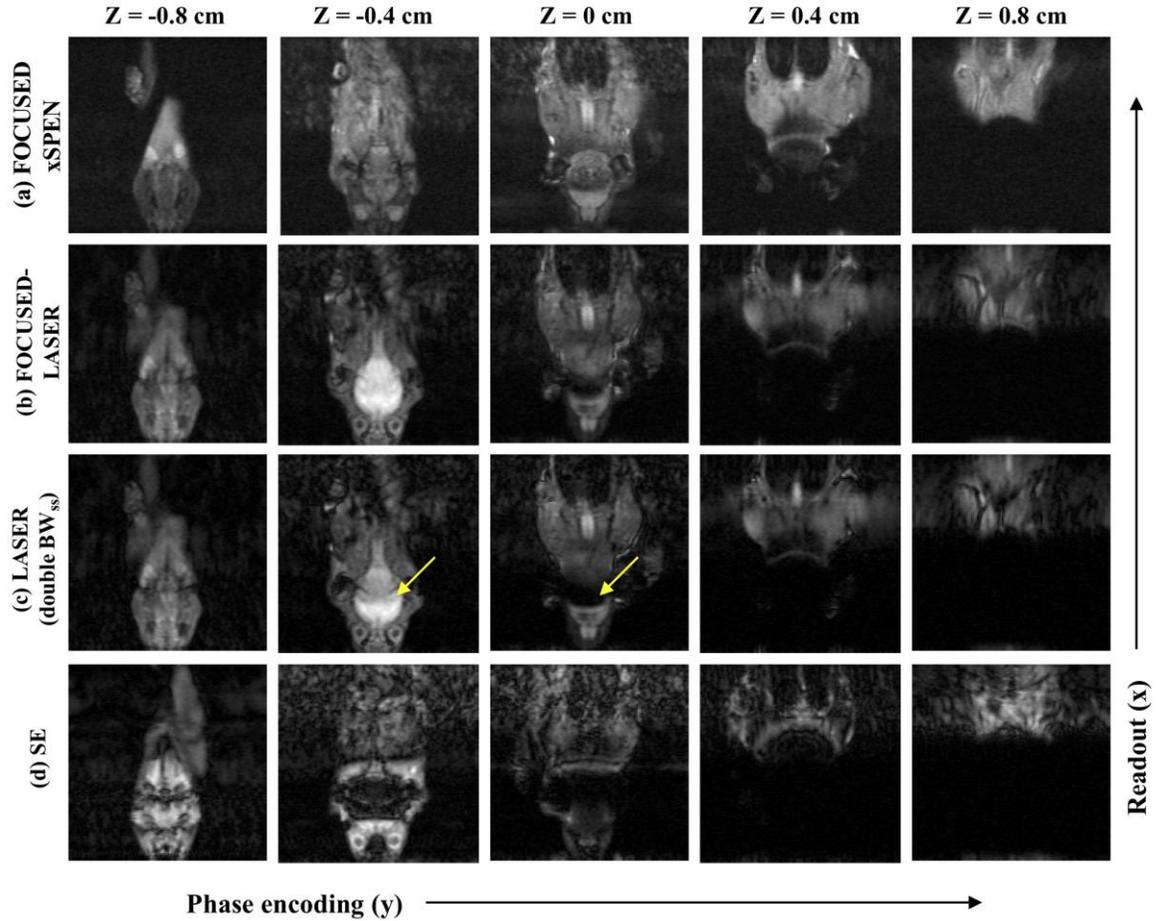

**Figure 7:** *In vivo* experiment performed on a similar setup as illustrated in Fig. 6, utilizing the indicated sequences. Swept pulses in FOCUSED-xSPEN (a), LASER (c) and FOCUSED-LASER (b) were 6 ms long and covered a bandwidth of 4 kHz (a) or 8 kHz (b-c), associated with a slice thickness of 4 mm. 4 ms long *sinc* pulses were used in the SE (d) experiment, effecting a bandwidth of 4 kHz. The LASER and FOCUSED-xSPEN started with a 0.25 ms hard pulse, exciting a bandwidth of ca. 4 kHz. In all methods involving looped transmission offsets, nine frequencies differing in 1 kHz steps were used. Other acquisition parameters were as used in Fig. 6. Yellow arrows highlight regions in the FOCUSED-LASER experiments with noticeable signal distortions. See Supporting Information Figure S9 for additional examples.

# New pulses for improving MRI's slice selectivity in the presence of strong, metal-derived inhomogeneities

Gil Farkash, Gilad Liberman, Ricardo P. Martinho and Lucio Frydman*

*Department of Chemical and Biological Physics, Weizmann Institute of Science, Rehovot 76100, Israel*

Figure 2a presented the amplitude profiles arising from implementing the four chirped pulses involved in the FOCUSED-xSPEN sequence. Figure S1 below shows how simulations predict the phases of the spins will accrue for each sweep of this sequence.

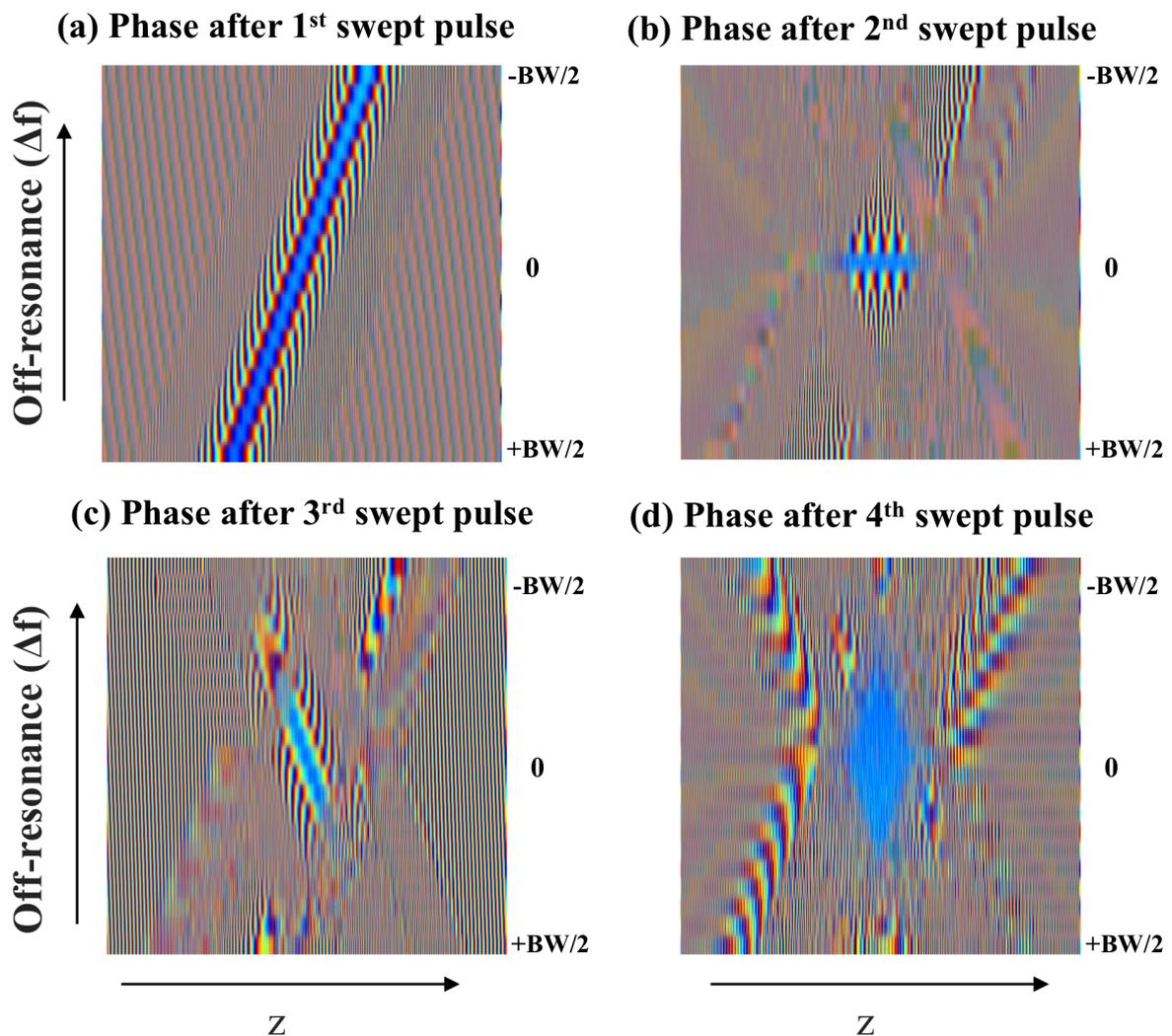

**Figure S1**: Bloch simulations showing how the phases of transverse magnetizations develop throughout the various stages of the FOCUSED-xSPEN sequence (left column of Fig. 1a). All RF pulses are assumed centered on-resonance; i.e., without looping the transmission offsets. The frequency swept pulses in this simulation swept through a bandwidth BW = 4 kHz associated with a region of 15 mm, while the targeted $z$ FOV was 30 mm long.

Figure 3 presented x-y images describing the performance of various sequences for a phantom including a Lego piece with a titanium screw. Figure S2 below zooms into the regions indicated by blue squares in the aforementioned images, to better illustrate the differences between the various methods.

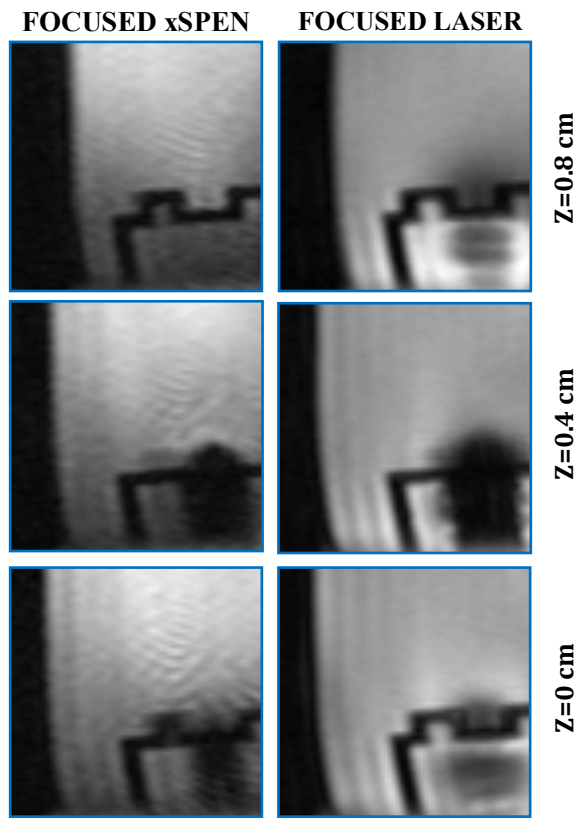

**Figure S2**: Zoomed regions arising from the $CuSO_4$-doped water tube containing a Lego piece and titanium screw introduced in Figure 3. The zoomed images were extracted from the full blue-line squares indicated in the original Figure.

The experimental results introduced in Figure 3 were acquired with transmission offset steps of BW/4 (1 kHz). Figure S3 presents additional experiments recorded on the same phantom, but with transmission offsets stepped in smaller increments: the FOCUSED-xSPEN and FOCUSED-LASER experiments had their center frequency steps incremented by BW/8 (500 Hz). The purpose of these repeat experiments was to figure out whether some of the 'banding' artifacts shown by the images in Figure 3, arose from poor overlapping among the spectral bands. Indeed, the experiments introduced in Figure S3 show smaller banding artifacts than those in Figure 3. SNR values were measured for these sequences relatively far from the metal-induced distortions (Figure S3, red square ROI) as the ratio between the ROI mean and its standard distortions, yielding: 13, 14, 16 and 38, , for the sequences in panels (a)-(d), respectively. Notice that the highest SNR values are then those arising from the SE sequence, probably as a result of being able to accommodate the shortest TE and thus be less exposed to diffusion/$T_2$ signal losses. Percent signal void areas were also calculated on the data in Figure S3 for the Z=0 slice, by counting the number of pixel values in the rectangular ROI indicated in yellow, that are smaller than the median value of all pixel values in the image. This classification between signal void and background areas was verified by reviewing all the corresponding binary masks, ensuring proper segmentation of the low-signal areas. The number of pixels that were less than the median was then normalized by the area of the ROI; the ensuing "lost pixel" fractions were 16%, 21%, 26% and 24%, for the images introduced in panels (a)-(d), respectively.

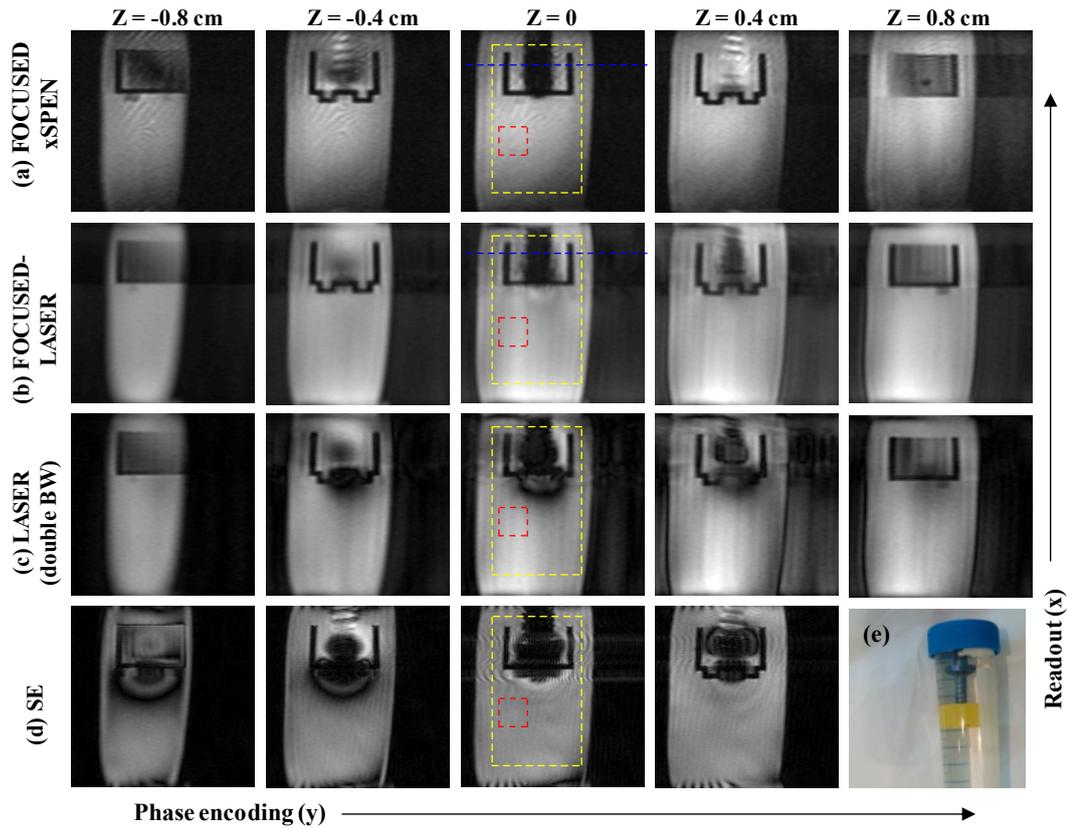

**Figure S3**: Images of the phantom from Figure 3 -depicted in (e)- acquired once again with a double- or four-times higher number of transmission offsets compared to Figure 3 to minimize 'band' artifacts associated with spectral overlap between spectral bins. Unlike in Figure 3, transmission offsets were shifted in steps of BW/8 or BW/16, increasing the overlaps between spectral bands. Uniformity values were measured in an ROI (red square) that is remote from the metal, yielding: 84% for FOCUSED-xSPEN, 87% and 86% for FOCUSED-LASER and LASER, and 91% for SE.

Figure S4 focuses on how the image sharpness and spatial resolution are preserved, by plotting 1D profiles extracted from the images introduced in Figure S3. The contrast of the Lego's piece slanted edge can be considered as a measure of spatial resolution, as degradations in the latter will also degrade contrast of edges in general. Image contrast in a region distant from the metal, was calculated as $Contrast = (I_{max} - I_{min})/(I_{max} + I_{min}) \cdot 100$ where $I_{max}$ and $I_{min}$ are the maximal and minimal image intensities in the segment between red bars.

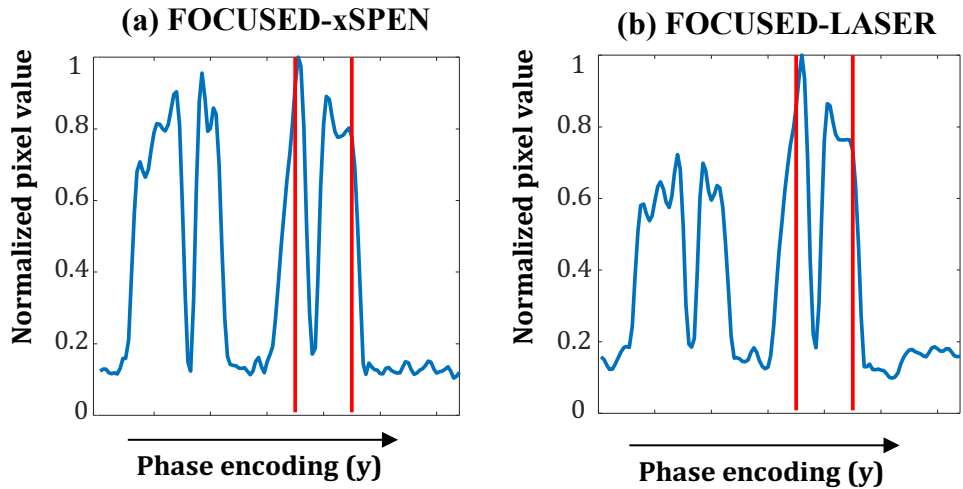

**Figure S4**: Normalized pixel value profiles of the rows marked by a blue dashed line in Figure S3. Contrast values were calculated in a region marked by red bars. This region contains an edge corresponding to the Lego piece. Image contrast in this region was calculated as $Contrast = 100\,(I_{max} - I_{min})/(I_{max} + I_{min})$, yielding 71% and 69% for (a) and (b) respectively.

Figure S5 presents another set of results collected for the same phantom, placed in a different position/orientation within the magnet.

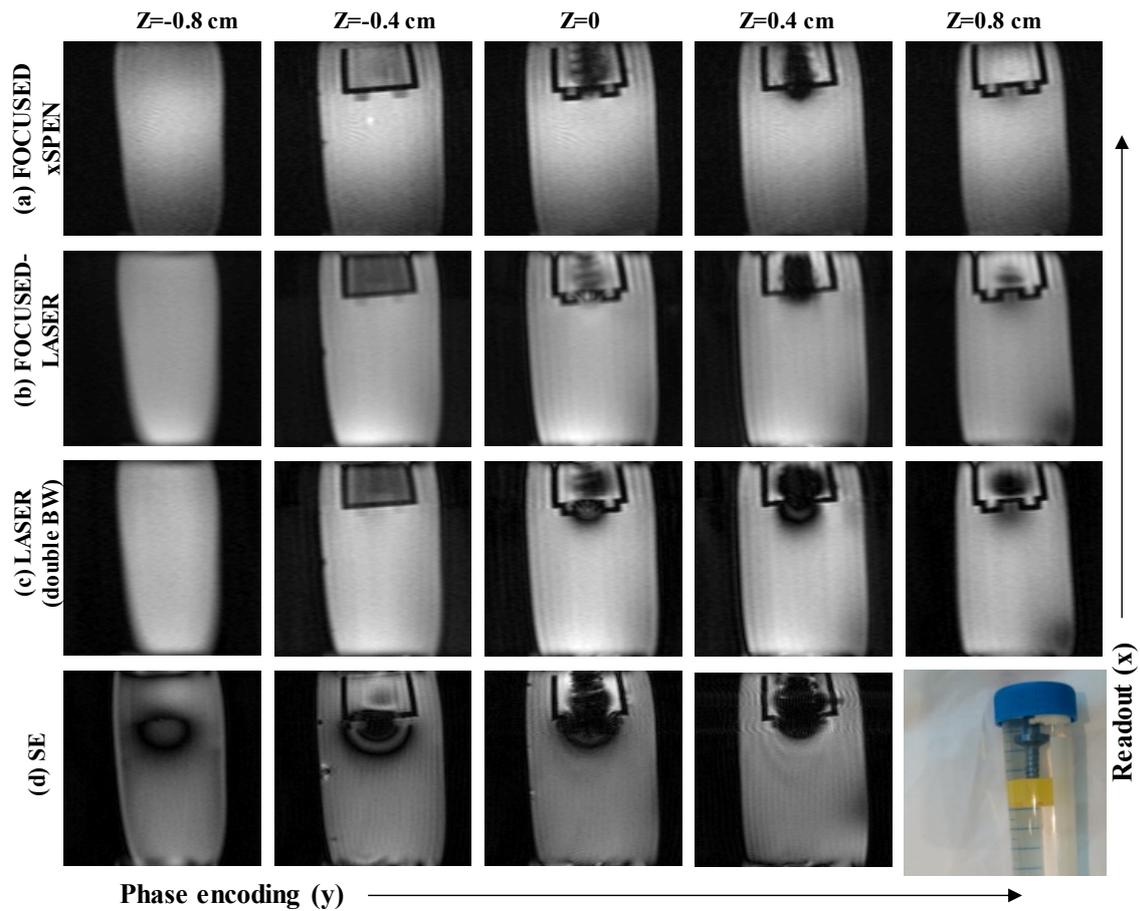

**Figure S5**: Additional *in vitro* experiments collected on the phantom used in Figures 3 and S3, acquired with the same pulse sequence and settings as in Figure 3 but for a different positioning of the phantom. Signal drops towards the end of the FOV in the FOCUSED-xSPEN acquisitions are probably reflecting RF inhomogeneity effects at the edge of the readout window or $T_2$ signal losses due to the longer echo times.

Figure 4 presented one comparison on the slice-selective accuracy of various sequences analyzed in this study. Figure S6 gives another perspective on this selectivity, by presenting the integrated projections of the various slices imaged for that titanium Lego phantom for each of the addressed slices. Yet another perspective regarding slice selectivity is presented in Figure S7, which instead of relying on imaging the slice-selection axis along the readout domain, was collected based on fully phase encoded variants of the various methods. This un-biased imaging block avoids possible distortion along the readout, which for some of the tackled inhomogeneities were as high as two pixels.

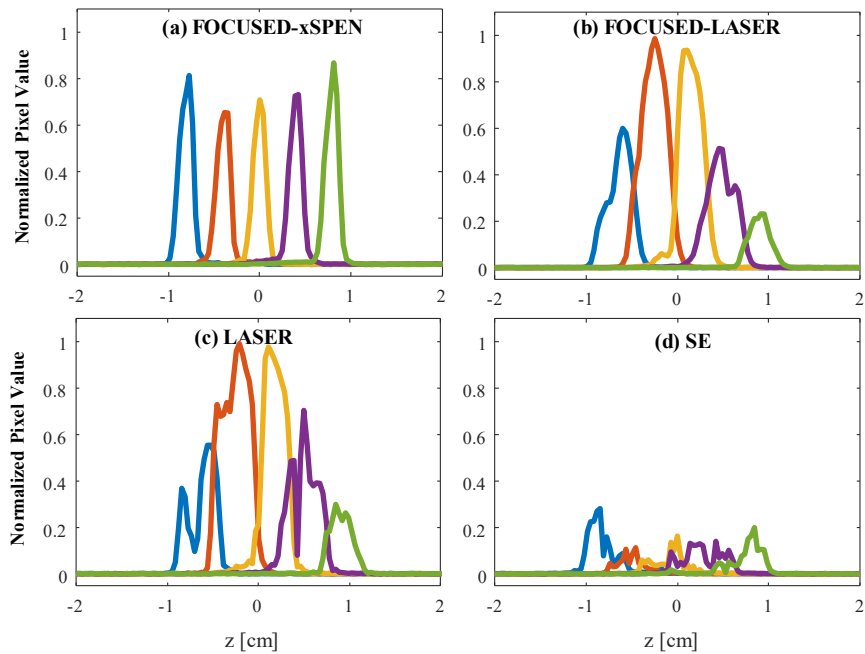

**Figure S6**: Slice profiles corresponding to the images in Fig. 4, as originated by the indicated sequences. The central columns from these multi-slice images are plotted as one-dimensional graphs, revealing inaccuracies in terms of the positions of the slice profiles as well as 'spill-over' effects in terms of their prescribed $z$ widths (4 mm in all cases). All plots are normalized to one being equal to the intensity of a slice with water, far away from the metal screw.

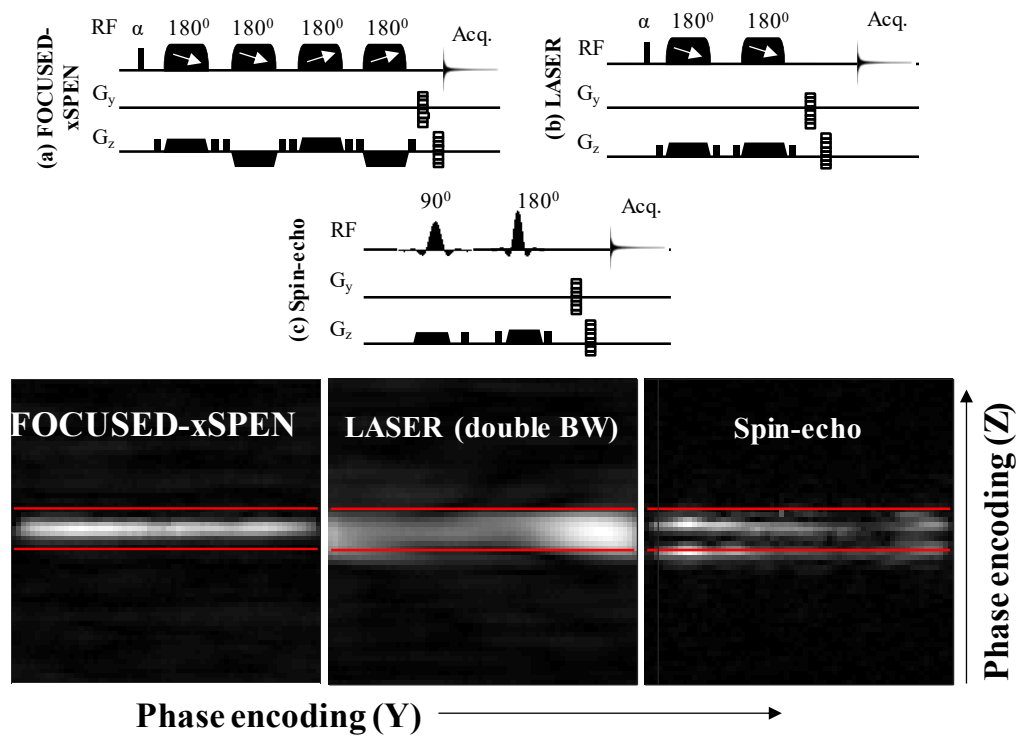

**Figure S7**: Fully phase encoded versions of the "Performance testing" sequences introduced in Fig. 1, for the sequences illustrated on top. The ensuing chemical shift imaging sequences thereby measure the accuracy of a slice selection along $z$, unbiased by inhomogeneity-derived shifts during the acquisition. A centered slice that was 4 mm wide (red lines) was targeted. The slice selection bandwidths for the spin-echo and FOCUSED-xSPEN experiments were 4 kHz, while LASER had a doubled slice selection bandwidth of 8 kHz. The acquisition was carried out with a spectral width of 4 kHz, and 512 time-samples for each phase encoded scan. A 64 x 64 matrix used to sample $k_z$ and $k_y$. Reconstruction was done by applying FT in all three dimensions ($k_z$-$k_y$-$t$) and summing all samples along the spectral dimension.

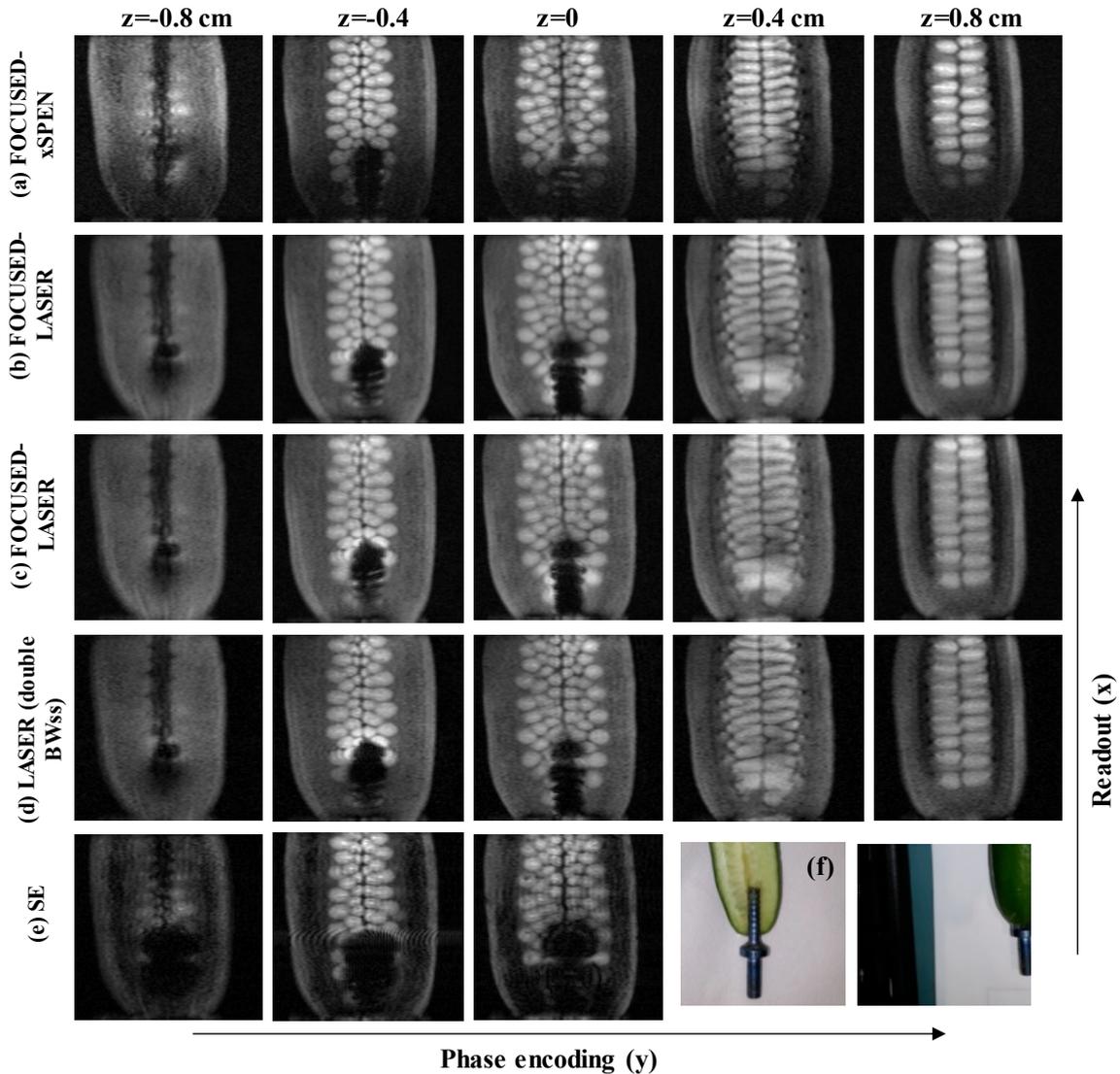

**Figure S8**: Similar scans as in Figure 5, but showing coronal images (instead of sagittals) for the phantom shown in panel (f). The slice thickness was 4 mm. FOCUSED-xSPEN swept adiabatic pulses had a bandwidth of 4 kHz while similar pulses in LASER/FOCUSED-LASER (b-d) had a doubled bandwidth (8 kHz). FOCUSED-LASER is shown in (b) with transmission offsets that impose Δf overlaps, while in (c) a subset of binned images was collected for the quadrature sum, thereby avoiding such overlap.

Figure 5 presented images describing the performance of various sequences in the excitation of sagittal slices for a phantom including a cucumber inserted with a titanium screw; Figure S8 complements these data with coronal slices acquired for a similar phantom.

Figures 6 and 7 presented images illustrating the performance of various slice-selection sequences in the excitation of images for a mouse with a titanium disk placed on its head; Figure S9 complements this with data acquired on a different animal on whose head a titanium cylinder was taped. This cylinder was characterized by comparable inhomogeneities to those arising from the disk. For instance, the FWHM in a pulse-acquire experiment that preceeded the imaging was 840 Hz with the titanium cylinder compared to 720 Hz with the titanium disk; 0.55% linewidths were ≈5.9 kHz for both titanium devices.

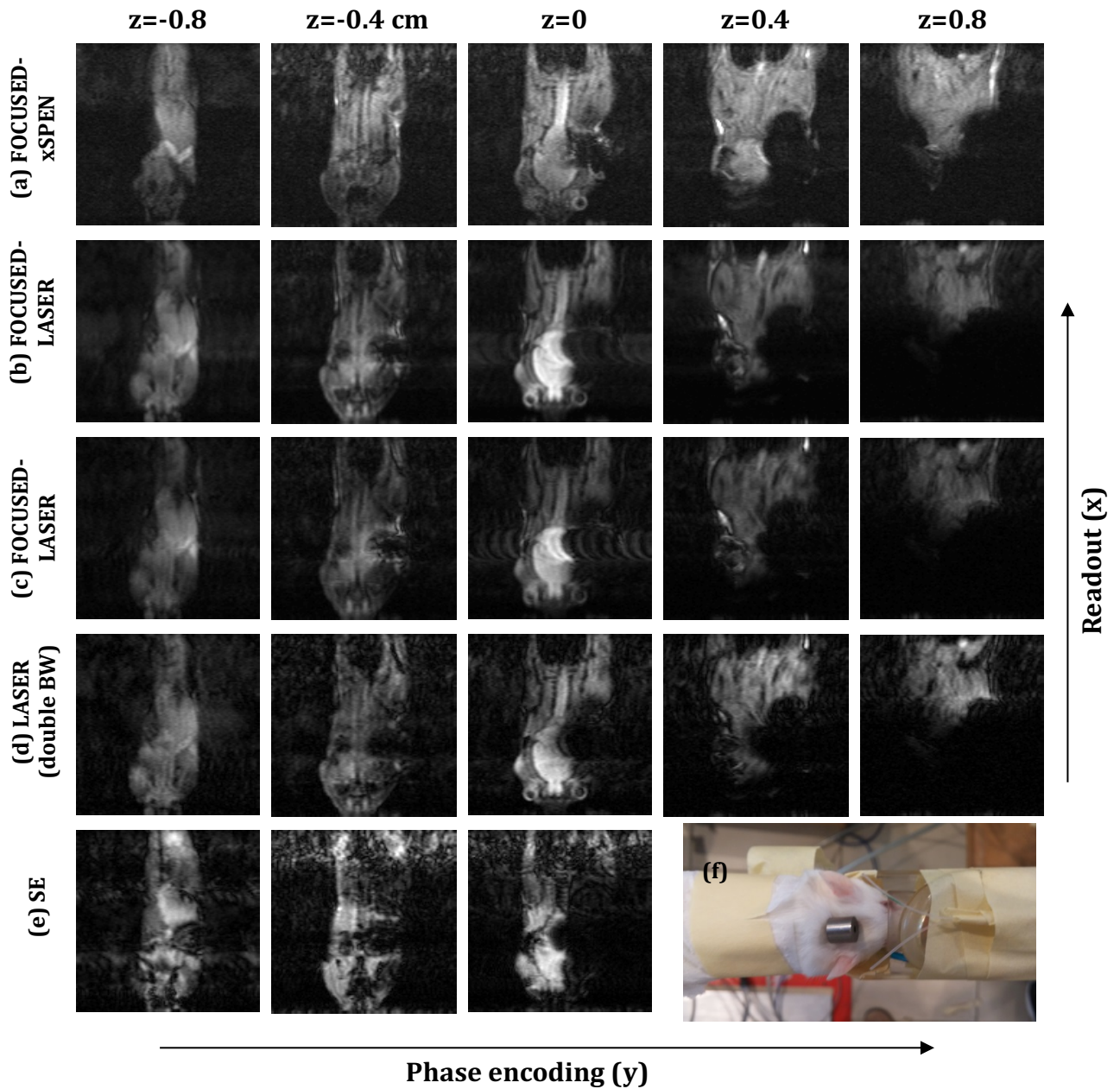

**Figure S9**: Mouse with titanium cylinder placed over its head (setup depicted in (f)), and scanned with the indicated pulse sequences. Acquisition parameters were as used in Fig. 7. Rows (b) and (c) display slices for FOCUSED-LASER experiments recorded with and without *Δf* overlaps, respectively.